\documentclass{optica-article}

\journal{opticajournal}
\articletype{Research Article}

\begin{document}

\title{Grammar-based topology search for reconfigurable integrated photonic spectrometers}

\author{Juejun Hu\authormark{1,*}}
\address{\authormark{1}Department of Materials Science and Engineering, Massachusetts Institute of Technology, 77
Massachusetts Avenue, Cambridge, MA 02139, USA}
\email{\authormark{*}hujuejun@mit.edu}

\begin{abstract*}
Reconfigurable integrated photonic spectrometers can generate exponentially increasing measurement states from a small number of switching elements, making them a promising design class for high-performance portable spectroscopy. However, selecting a circuit architecture remains largely intuition driven. We introduce a grammar-based topology optimization framework that composes standard photonic building blocks, canonicalizes and physically prunes candidate circuits, and jointly searches topology and component parameters using a coherent scattering-matrix model. Designs are screened by a decoder-independent architecture objective that combines absolute singular-value noise gain with multi-line conditioning, thereby accounting for response diversity and optical throughput. Direct held-out spectrum reconstruction confirms that this surrogate is a strong predictor of reconstruction quality at high incident signal-to-noise ratio (SNR). Across the nonresonant and resonator-augmented design spaces examined, dFT and its variants offer the best balanced performance; increased component loss favors a shallower dFT variant in which one differential-delay stage is replaced by an analog optical phase shifter. Passive and active rings provide no robust advantage attributable to resonant amplitude coding. The results identify balanced complementary interferometric responses, efficient terminal collection, and limited lossy circuit depth as central design rules, and establish a systematic route for selecting task-optimized photonic circuit topologies under technology-specific constraints.
\end{abstract*}

\section{Introduction}

Optical spectroscopy is indispensable to chemical and materials analysis, environmental and biological sensing, hyperspectral imaging, and optical-network monitoring. Conventional benchtop instruments can deliver exceptional spectral resolution and bandwidth, but their bulk, cost, alignment requirements, and, in some cases, moving parts limit deployment outside controlled laboratory settings. Photonic integration offers a route to compact, mechanically robust, and potentially manufacturable spectrometers that can bring high-quality spectral analysis to portable, distributed, and in situ applications~\cite{Yang2021Miniaturization,Li2022Advances,Peters2025CriticalReview,Zhang2025IntegratedReview}. This opportunity has motivated rapid development of integrated spectrometers based on dispersive elements~\cite{Cheben2007AWG}, resonator arrays~\cite{Gan2012CavityArray}, Fourier-transform interferometers~\cite{Velasco2013FTS}, and engineered random scattering media~\cite{Redding2013Disordered}, among other approaches~\cite{LeCoarer2007SWIFTS,Kyotoku2010Cavity,Bock2013Subwavelength,Redding2016Spiral,Nedeljkovic2016MidInfrared,Nie2017Stationary,Piels2017Multimode,Madi2018Lippmann,Hartmann2020Disorder,Pohl2020LithiumNiobate,Li2021Stratified,Zhang2021Tandem,Zhang2021Speckle,Lin2023Speckle}.

Reconfigurable photonic integrated circuits are particularly attractive for optical spectroscopy because one physical circuit can be programmed to provide many distinct spectral measurements. A circuit containing $K$ independent control variables, each supporting $N$ states, can access as many as $N^K$ configurations; more generally, if control variable $k$ supports $N_k$ states, the state count is $M=\prod_{k=1}^{K}N_k$. Multiple physical components may share one control variable, and additional routing switches need not add independent states. This exponential scaling allows the number of calibrated measurement channels to grow rapidly without requiring a corresponding number of replicated optical outputs or detectors. The bandwidth-to-resolution ratio, which approximates the number of resolvable spectral intervals within the operating band, is widely reported as a spectrometer figure of merit~\cite{Yao2023Broadband,Yao2024Metrology,Xu2023TemporalSpeckle}. Switch-based architectures are especially attractive in this regard: their exponentially growing state space offers a route to exceptionally high bandwidth-to-resolution ratios without proportional growth in optical hardware footprint, component count or complexity~\cite{Kita2017Scaling,Kita2018DFT,Du2022OpticalSwitches}. For example, the digital Fourier-transform (dFT) spectrometer uses binary optical switches that select combinations of path delays in a programmable interferometer~\cite{Kita2017Scaling,Soref2018Digital,Kita2018DFT,Pavanello2019Digital,Du2022OpticalSwitches,Wei2023MidInfrared,Mojahed2024OCTDFT,NavaBlanco2025Glucose,NavaBlanco2025SkinCancer,Peters2026BroadbandDFT}. More broadly, reconfigurable spectrometers have been implemented with thermo- or electro-optically tuned interferometers~\cite{Souza2018ThermoOptic,Zheng2019RingAssisted,Montesinos2019ThermoOptic,Li2020FabricationTolerant,Li2021IntegratedFTS,Xu2024TwoDimensional,Cui2026CascadedMZI}, MEMS-actuated interferometers and couplers~\cite{Fathy2020Parallel,Qiao2022MEMS,Chen2024MEMS,Sun2024Digitalized,Zhou2024Denoising}, switch-selected delay networks~\cite{Soref2018Digital,Pavanello2019Digital,Du2022OpticalSwitches}, tunable resonator filters~\cite{Zheng2019TunableMicroring,Sun2022FSRFree,Zhang2022Scanning,Xu2023PhotonicMolecule,Xu2023SingleResonator,Sun2023Microdisk,Zhang2025ResolutionSwitchable,Liu2025MicroringNetwork}, and reconfigurable multimode or programmable networks~\cite{Yi2021PhotonicLantern,Yi2022TwoStage,Xu2023TemporalSpeckle,Yao2023Broadband,Yao2023Programmable,Li2024Reconfigurable,Yao2024Metrology}. Nevertheless, exponential growth in state count does not by itself guarantee an equal number of useful spectral degrees of freedom: insertion loss, redundant or weakly distinguishable responses, finite calibration resolution, and noise amplification during reconstruction can all erode the nominal scaling advantage.

The expanding range of proposed architectures therefore raises a simple but consequential question: \emph{which photonic circuit topology yields the best spectral reconstruction quality?} Addressing this question, however, requires solving two coupled problems. First, circuit topology must be expressed in a rigorous mathematical language that is broad enough to represent disparate architectures, constrained enough to exclude unphysical candidates, and amenable to automated search. Second, candidate topologies must be evaluated using a performance measure that reflects spectrum reconstruction under realistic noise rather than an isolated optical proxy. These challenges are inseparable because topology determines how optical power is divided, delayed, mixed, routed, and detected, thereby controlling both response diversity and signal throughput.

Automated photonic design has begun to address topology as well as continuous device parameters. Graph-based mutation has been used to evolve complete photonic systems~\cite{MacLellan2024InverseSystems}; graph-theoretic searches have explored microwave-photonic architectures and coupled-mode scattering setups~\cite{Li2025AutomaticMWP,Landgraf2025AutoScatter}; differentiable and evolutionary searches have identified photonic tensor-core circuits subject to hardware constraints~\cite{Gu2022ADEPT,Jiang2025ADEPTZ}; and graph-based algorithms have designed and simplified quantum optical experiments~\cite{Krenn2021Automated}. While these advances establish the value of automated architecture discovery, the need for a circuit-level formulation that can systematically compose standard photonic building blocks, enforce physical constraints before simulation, recover known spectrometer families, and compare them under a reconstruction-specific objective remains largely unaddressed.

The second challenge is not met by traditional scalar resolution proxies alone. The half-width at half-maximum of the spectral correlation function is often reported as an estimate of resolution in reconstructive spectrometers~\cite{Redding2013Disordered,Redding2016Spiral,Hartmann2020Disorder,Yao2023Broadband,Yao2023Programmable,Cui2026CascadedMZI}. Optimizing only this width, however, can favor response families whose correlation functions have narrow central lobes but large oscillatory sidelobes; these secondary similarities create spectral ambiguities and increase sensitivity to noise. Moreover, rescaling every response by the same attenuation factor leaves the normalized correlation function and its half-width unchanged even though reconstruction becomes more susceptible to signal-independent detector noise. Pairwise correlation statistics also do not fully describe the conditioning of simultaneous multi-line reconstruction. A recently proposed inference-oriented objective instead uses the nuclear norm of the measurement-matrix pseudoinverse to account jointly for response distinguishability and collection efficiency~\cite{Ma2026RobustInference}. This singular-value perspective provides a natural foundation for comparing circuit architectures under a common reconstruction model.

This work makes three principal technical contributions. First, we introduce, to our knowledge, the first grammar-based circuit-topology optimization framework for reconfigurable integrated photonic spectrometers. Unlike approaches that optimize component parameters within a prescribed architecture, the framework searches circuit connectivity by composing standard photonic building blocks, canonicalizing equivalent descriptions, and pruning physically inactive or redundant candidates before numerical evaluation. We combine this discrete search with continuous parameter optimization using an S-matrix circuit model. Although demonstrated here for spectrometers, the grammar-based formulation is not specific to spectroscopy: its component alphabet, port constraints, physical rules, and objective can be adapted to search circuit architectures for a broad range of photonic integrated circuit (PIC) applications.

Second, building on the singular-value perspective of Ref.~\cite{Ma2026RobustInference}, we propose a simple scalar, decoder-independent metric, $J$, for quantitatively comparing reconstructive spectrometers. It combines absolute reciprocal-singular-value noise gain---equivalent, up to normalization and rank conventions, to the nuclear norm of the measurement-matrix pseudoinverse---with the conditioning of equal-strength three-line spectra. The metric therefore evaluates response diversity, optical throughput, and susceptibility to signal-independent detector noise without assuming a particular reconstruction algorithm. Because it is defined from the calibrated measurement matrix rather than any specific photonic implementation, it is broadly applicable to reconstructive spectrometers, including static wavelength multiplexing systems implemented using either integrated photonic platforms or conventional bulk optical elements. We validate $J$ against direct held-out spectrum reconstruction across a broad topology cohort, multiple incident SNRs, and representative decoders, while retaining the distinction between an architecture-screening metric and decoder- and task-dependent reconstruction error.

Third, we apply the framework to integrated photonic spectrometers reconfigured through two widely adopted switching elements---waveguide phase shifters and binary optical switches---and identify dFT and its variants as the highest-performing topology family within the design spaces evaluated. The search further yields physics-based design rules that explain this result in terms of balanced interferometric coding, optical throughput, component loss, and reconstruction conditioning, rather than presenting only a black-box ranking. These contributions collectively provide general tools for PIC architecture discovery and reconstructive-spectrometer assessment, while establishing a reproducible basis for selecting and rationally improving reconfigurable integrated photonic spectrometer design under technology-specific constraints.

\section{Methods}

\subsection{Design scope and grammar-based topology representation}

\begin{figure*}[t]
\centering
\includegraphics[width=\linewidth,height=0.78\textheight,keepaspectratio]{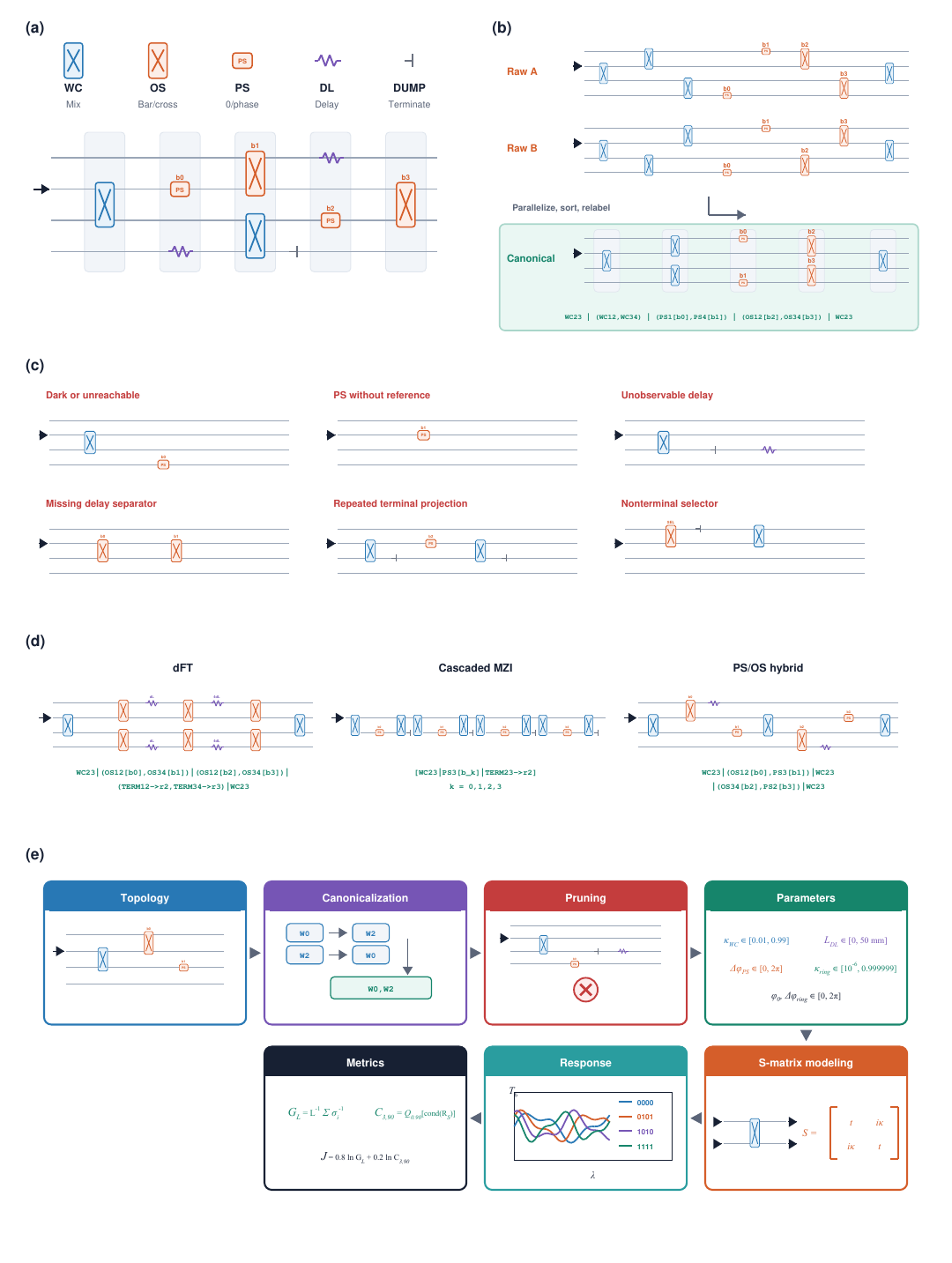}
\caption{Grammar-based circuit-topology search. (a) WCs, binary OSs and PSs, DLs, and dumps are composed into nonoverlapping layers on four rails. Blue, orange, and purple denote passive, active (binary switching), and delay elements, respectively. (b) Circuits differing only by commuting operations or bit names converge to one canonical diagram and word. (c) Pruning removes dark, unreachable, phase-insensitive, or exactly redundant structures. (d) The grammar naturally recovers a dFT, a cascaded MZI filter, and a representative PS/OS hybrid; the corresponding words are shown beneath the diagrams. (e) The modeling workflow: candidate topologies are canonicalized and pruned before continuous-parameter optimization, S-matrix propagation, response calculation, and evaluation of $G_L$, $C_{3,90}$, and $J$.}
\label{fig:method_overview}
\end{figure*}

We represented a spectrometer as an ordered sequence of component layers on four parallel single-mode waveguide rails [Fig.~\ref{fig:method_overview}(a)]. Light was launched into rail 2, and the optical powers at the two central output rails (rails 2 and 3) were detected independently. This dual-output configuration is analogous to complementary-beam detection in classical Fourier-transform spectrometers: when the two detector noises are independent and dominant, combining the complementary interferograms can improve the signal-to-noise ratio by a factor of $\sqrt{2}$ relative to using one output~\cite{Harwit1979Hadamard}. The component alphabet comprised wavelength-insensitive waveguide couplers (WCs), binary optical switches (OSs), binary phase shifters (PSs), delay lines (DLs), and terminal-routing elements. A terminal-routing element, which can practically be implemented as a $2\times2$ optical switch, is an end-of-branch two-rail site that directs the state-dependent retained field toward detection and explicitly terminates the unused rail; its route is determined from the existing control bits and does not add an independent state. A WC coherently mixed adjacent rails with a continuously optimized power-coupling coefficient. An OS selected either its bar or cross connection, whereas a PS applied either zero phase or a continuously optimized phase swing. A DL introduced an optimized wavelength-dependent propagation phase and the corresponding propagation loss. The binary control variables were shared across all components assigned to the same bit, so a circuit with $K$ independent bits produced $2^K$ optical states even when its total number of physical OSs and PSs exceeded $K$. The principal study used $K=4$, giving 16 states and 32 detector responses; the selected canonical dFT contained four independently controlled OSs plus two terminal-routing OSs. A $K=6$ study used 64 states and 128 detector responses.

Each circuit was serialized as a grammar word read from the input toward the detectors. For example,
\[
\begin{aligned}
&\mathrm{WC}_{23}\;|\;(\mathrm{WC}_{12},\mathrm{WC}_{34})\;|\;
(\mathrm{PS}_{1}[b_0],\mathrm{PS}_{4}[b_1])\;|\\
&\hspace{4em}(\mathrm{OS}_{12}[b_2],\mathrm{OS}_{34}[b_3])\;|\;\mathrm{WC}_{23}
\end{aligned}
\]
denotes five successive component layers separated by vertical bars. Components within parentheses occupy the same layer and act in parallel; component subscripts identify the affected rail or adjacent rail pair, and $[b_k]$ identifies the binary control bit.

The circuit grammar generated feed-forward interconnection graphs: after an optical field passed a component layer, no circuit-level connection returned it to an earlier layer. The first layer was a required WC between the two central rails, ensuring that the injected field could access both detected paths. Subsequent layers contained non-overlapping one- or two-rail elements. For $K=4$, every candidate contained exactly four independent active OS or PS bits, and WCs and generic terminal-routing sites were counted together under an eight-site connectivity limit. This limit is sufficient to represent all of the common nonresonant topology classes investigated here, including concatenated switch-selected delay networks such as dFT~\cite{Kita2017Scaling,Kita2018DFT,Du2022OpticalSwitches}, coherent cascades of phase-tunable interferometers~\cite{Cui2026CascadedMZI}, and cascaded MZI filters~\cite{Yao2023Programmable}. DL variables could be placed only on optically active rail intervals between two-rail sites. A generic terminal site was instantiated as a state-programmed OS that selected the desired output rail, followed by an explicit dump of the unused output. For each of the $2^K$ states, its bar/cross setting was the discrete inner choice that routed the larger wavelength-averaged input power to the retained rail; the resulting truth table was frozen for final evaluation and exact replay. This routing used the existing $K$ bits, created no additional state dimension, and is reported explicitly in the Supplementary Information. A WC followed by a dump was retained as an alternative terminal implementation for control audits, as in an MZI terminated with a $2\times1$ combiner~\cite{Yao2023Programmable}.

The feed-forward restriction applies to the circuit interconnection graph. Local resonant components were admitted in the ring studies only as fixed-port scattering blocks within bounded augmentation and replacement catalogues, even though their transfer functions contain internal recirculation. Spectrometers whose overall operating principle relies on a non-feed-forward cavity or recirculating network~\cite{Xu2023TemporalSpeckle} are outside the scope of the present topology search. The grammar likewise did not enumerate nonadjacent couplers, multimode devices, arbitrary resonator networks, or unbounded passive complexity.

Candidate circuits were pruned before numerical optimization [Fig.~\ref{fig:method_overview}(c)] when: 1) a component lay outside every illuminated source-to-detector path, 2) a PS lacked an unaffected coherent reference at either detector, 3) a DL lay on an unreachable or phase-invisible interval, 4) a reducible same-pair component sequence lacked a required differential-delay separator, 5) a terminal rank-one projection on a rail pair was repeated before a two-rail element could remix that pair, or 6) a terminal selector was followed by an operation that reused its terminated outer rail. Single-rail PS and DL operations between two same-pair terminal projections cannot restore the discarded optical degree of freedom, so the second projection has an exactly equivalent simpler representation. As illustrated in Fig.~\ref{fig:method_overview}(b), after normal-form ordering, bit identifiers were reassigned sequentially from $b_0$ in order of first appearance, after which the ordered component tokens and their rail supports formed the topology key; duplicate keys were retained only once. Mirror equivalence was applied only when it preserved the fixed input and detector ports. These conservative rules removed candidates that were structurally inactive or exactly redundant without ranking circuits by a heuristic performance estimate.

As shown in Fig.~\ref{fig:method_overview}(d), topologies produced by the grammar-based search scheme naturally encompass known architectures such as the dFT with state-programmed terminal routing~\cite{Kita2019ChipScaleDFT}, coherent phase-chain architectures~\cite{Cui2026CascadedMZI}, and cascaded Mach--Zehnder interferometers with terminated outputs~\cite{Yao2023Programmable}. Their continuous parameters were optimized under the same bounds and objective as the generated circuits. The $K=6$ grammar extended the independent-bit count to six while retaining the same four-rail input--output definition.

\subsection{Scattering-matrix circuit model}

The topology-response-metric evaluation pipeline is summarized in Fig.~\ref{fig:method_overview}(e). Each candidate was evaluated using a coherent frequency-domain scattering-matrix model. For every wavelength and binary state, a four-component complex field vector was initialized with unit amplitude on rail 2 and propagated through the ordered component list. An adjacent-rail WC with power-coupling coefficient $\kappa$ was represented by
\begin{equation}
\mathbf{S}_{\mathrm{WC}}=a_{\mathrm{WC}}
\begin{pmatrix}
\sqrt{1-\kappa} & i\sqrt{\kappa}\\
i\sqrt{\kappa} & \sqrt{1-\kappa}
\end{pmatrix},
\end{equation}
where $a_{\mathrm{WC}}$ is the field transmission associated with its insertion loss. An OS applied either the identity (bar) or exchange (cross) matrix, multiplied by its field transmission. A PS on rail $r$ multiplied that field by $\exp(i b_k\phi)$, where $b_k\in\{0,1\}$ is the corresponding state bit and $\phi$ is the optimized binary phase swing. A DL of physical length $l_{\mu\mathrm{m}}$ applied
\begin{equation}
t_{\mathrm{DL}}(\lambda,l)=10^{-\alpha_{\mathrm{wg}}l_{\mathrm{cm}}/20}\exp[i\beta(\lambda)l_{\mu\mathrm{m}}],
\end{equation}
where $l_{\mathrm{cm}}=l_{\mu\mathrm{m}}/10^4$, $\alpha_{\mathrm{wg}}$ is in dB/cm, and $\beta$ is the propagation constant in rad/$\mu$m. The propagation constant was calculated from the effective and group indices and, in the dispersion studies, the run-specific waveguide, PS, and OS dispersion models; the full wavelength-dependent expressions are given in Supplementary Sec.~S1. Output powers were obtained as the squared field magnitudes on the two detector rails and stacked over states to form a nonnegative response matrix $\mathbf{R}$.

Unless otherwise stated, simulations covered 1530--1565~nm, used an effective index of 2.5 and a group index of 4.2 at 1550~nm, and assigned losses of 1~dB/cm to explicitly represented DL length, 0.2~dB per WC, and 0.4~dB per OS in either state. These values were chosen as representative foundry-scale SOI parameters~\cite{Fahrenkopf2019AIM}. PSs, bends, and waveguide transitions were assumed to be lossless. The doubled-loss study used 2~dB/cm, 0.4~dB per WC, and 0.8~dB per OS, while retaining the assumption of zero PS insertion loss. The expanded-band dispersion study covered 1500--1600~nm. WC coupling was assumed to be wavelength independent, emulating the response of adiabatic broadband couplers~\cite{Yun2013Adiabatic}. These idealized component models isolate topology-level tradeoffs; fabrication error, thermal crosstalk, polarization dependence, nonlinearity, detector correlation, and experimental model mismatch were not included.

Passive and active rings were modeled by their complex through-port or add--drop scattering responses and inserted as local blocks in the feed-forward circuit. The passive-ring catalogue optimized free spectral range (2--20~nm), intrinsic quality factor ($10^2$--$10^8$), bus power coupling ($10^{-6}$--0.999999), and resonance offset ($0$--$2\pi$). Active rings used the same ranges and added a binary tuning phase from $0$ to $2\pi$. These studies compare bounded ring augmentations or replacements with ring-free references; they are not exhaustive searches over resonator-network topology.

\subsection{Architecture-screening objective and direct reconstruction validation}

We screened circuits using a noise-aware objective defined from the calibrated measurement response rather than from the width of a spectral-correlation peak. Let $\{\psi_l(\lambda)\}_{l=1}^{L}$ be a shifted Chebyshev basis over the analysis band, orthonormalized using trapezoidal quadrature weights $\mathbf{W}$. The calibrated response matrix $\mathbf{R}_{\mathrm{cal}}$ was obtained by convolving each monochromatic detector response with the Gaussian calibration-source line and then cropping a four-standard-deviation simulation guard band; it maps a sampled input spectrum to the detector measurements across all switching states. The corresponding coefficient-to-measurement map was
\begin{equation}
\mathbf{A}=\frac{1}{\sqrt{2^K}}\mathbf{R}_{\mathrm{cal}}\mathbf{W}\boldsymbol{\Psi},
\end{equation}
where the columns of $\boldsymbol{\Psi}$ contain the $L$ basis functions sampled on the wavelength grid. The factor $1/\sqrt{2^K}$ models a fixed total acquisition time divided equally among sequential switch states; the two detector outputs within each state were treated as simultaneous measurements. We assumed equal, independent, signal-independent detector noise after this time normalization.

If $\sigma_l$ are the absolute singular values of $\mathbf{A}$, the first objective component was
\begin{equation}
G_L=\frac{1}{L}\sum_{l=1}^{L}\frac{1}{\sigma_l}.
\end{equation}
This is the mean singular-direction noise gain and is proportional to the nuclear norm of the measurement-matrix pseudoinverse~\cite{Ma2026RobustInference}. Because the response amplitudes were not normalized away before computing $\mathbf{A}$, $G_L$ jointly penalizes poorly encoded spectral modes and noise amplification caused by low optical throughput. We used $L=12$ for $K=4$ and $L=48$ for $K=6$, prespecified smooth-spectrum basis sizes that increased with the available response dimension while remaining below the corresponding 32 and 128 detector responses.

The second component tested the stability of sparse multi-line reconstruction. Columns of $\mathbf{R}_{\mathrm{cal}}$ were normalized to unit Euclidean norm, and condition numbers were calculated for sampled three-column submatrices representing equal-strength three-line spectra. Neighboring wavelengths in each sampled three-line set were separated by at least $B/L$, where $B$ is the analysis bandwidth. We denote the empirical 90th percentile of these condition numbers by $C_{3,90}$. This statistic detects multi-column aliases that can be missed by a correlation-function half-width or a single pairwise coherence value.

The scalar architecture-screening objective was
\begin{equation}
J=0.8\ln G_L+0.2\ln C_{3,90},
\end{equation}
with lower values preferred. The weights emphasize broadband noise amplification while retaining a conservative test of sparse spectral ambiguity. A sensitivity analysis over the fixed 115-design validation cohort found that the selected weight lies on a broad near-optimal plateau; blind three-line reconstruction and topology-group and stage-2 holdouts further showed no reliable advantage from fitting the weight or selecting a more flexible two-term family. These results support the selected form of $J$ (Supplementary Sec.~S9). Mean and minimum throughput, pairwise coherence, other singular-value orders, state separation, bit influence, calibration visibility, correlation functions, Cram\'{e}r--Rao bounds, and two-line discrimination were retained as diagnostics but did not enter $J$; these diagnostics are provided in the Supplementary Information. Because $L$, state count, response dimension, and the $B/L$ three-line separation rule change between studies, numerical $J$ values were compared only among candidates evaluated within the same campaign; in particular, C-band and expanded-band scores, and $K=4$ and $K=6$ scores, are not directly comparable. We also note that $J$ is a decoder-independent architecture-screening surrogate, not direct reconstruction error: it does not account for decoder regularization, task priors, or finite-SNR nonlinearities.

We therefore performed a separate direct reconstruction validation under additive, independent Gaussian detector noise. Every incident spectrum was normalized to unit quadrature-weighted power, and the normalized incident SNR was defined as $\mathrm{SNR}_{\mathrm{in}}=1/\sigma_{\mathrm{det}}$, where $\sigma_{\mathrm{det}}$ is the standard deviation of the detector noise in those units; its decibel value is $20\log_{10}(\mathrm{SNR}_{\mathrm{in}})$. To compare different state counts at fixed total acquisition time, the calibrated response for each sequential state was divided by $\sqrt{2^K}$; the two detector outputs in a state were acquired simultaneously. A 200-design $K=4$ cohort combined 100 densely evaluated finalists with 100 deterministic stage-2 candidates stratified over objective score ranges and structural signatures. For each design, $J$ was recomputed on a common grid and compared at $\mathrm{SNR}_{\mathrm{in}}=10$, 30, 100, and 300 with the empirical composite error
\begin{equation}
E_{\mathrm{rec}}=0.8\ln \widetilde{\epsilon}_{\mathrm{broad}}+0.2\ln \epsilon_{\mathrm{sparse},90},
\end{equation}
where $\widetilde{\epsilon}_{\mathrm{broad}}$ is the median normalized root mean square error (NRMSE) from 128 held-out positive broadband spectra with four shared-noise repeats, and $\epsilon_{\mathrm{sparse},90}$ is the 90th-percentile error for 64 separated equal-strength three-line wavelength sets with four repeats. The broadband decoder was ridge regression on the same 12-term quadrature-orthonormal Chebyshev basis, with regularization selected from 64 validation spectra separately for every design and SNR. For the equal-strength three-line wavelength sets, the true three-line wavelengths were supplied to the decoder, which estimated only their amplitudes. This removes errors from locating the spectral lines and isolates how reliably the response matrix distinguishes and quantifies a known three-line spectrum. The rank analysis in the main text uses the 115 competitive designs with recomputed $J<5$; the complete 200-design range, including near-rank-deficient candidates, is reported in the Supplementary Information. Spearman confidence intervals were obtained from 1,000 bootstrap resamples within the stated analysis set.

We additionally tested five representative $K=4$ architectures with nonnegative decoders and denser preserved response matrices. The broadband task used 48 validation spectra to tune second-difference Tikhonov regularization and 96 held-out smooth or mixed spectra, each with four shared-noise repeats; NRMSE was evaluated in the quadrature-weighted spectral norm. The sparse task used 72 validation and 144 held-out one-, two-, and three-line spectra of equal or unequal strengths, again with four repeats, on a 64-point wavelength grid. A nonnegative $\ell_1$ decoder was tuned independently for every design and SNR, and sparse NRMSE and mean line localization error were retained as diagnostics. All designs received identical truth spectra and detector noise realizations within each task.

\subsection{Discrete--continuous search and computational provenance}

For the nonresonant topology campaigns, a stratified grammar sampler produced a reservoir of 100,000 distinct topologies. Continuous WC power couplings were bounded between 0.01 and 0.99, PS phase swings between 0 and $2\pi$, and each physical DL length between 0 and 50,000~$\mu$m. To sample short and long delays efficiently, the optimizer used the coordinate $x=\log_{10}(1+l/\mu\mathrm{m})$, corresponding to the physical length $l=(10^x-1)\,\mu\mathrm{m}$. The corresponding 50-mm length bound was only a numerical sampling ceiling. Before evaluation in the $K=4$ campaigns, if the conservative sum of proposed DL lengths exceeded the optical path difference compatible with less than 5\% fringe-visibility loss for the finite-linewidth calibration source, all DL lengths were reduced proportionally; candidates that still failed the calibrated-response check were rejected. The $K=6$ campaign applied the analogous limit to the actual state-resolved coherent path span. Continuous candidates combined scrambled Sobol samples, multiscale local perturbations around surviving designs, and bounded derivative-free Powell polishing. Fixed seeds made the sampled topology reservoirs and continuous proposals reproducible.

The nominal $K=4$ search used four successive wavelength-grid spacings of 0.25, 0.10, 0.05, and 0.025~nm and evaluated 32, 64, 128, and 256 separated three-line wavelength sets, respectively. Successive stages retained 10,000, 1,000, and 100 candidates before final polishing. The planned evaluation budget was $10^6$ topology--parameter evaluations, allocated as 400,000, 200,000, 200,000, and 200,000 evaluations across the four stages. Three named reference architectures were carried through every shortlist independently of their preliminary score. Campaign-specific changes for doubled loss, dispersion, expanded bandwidth, $K=6$, and ring catalogues were fixed before their corresponding searches and are reported with the archived run specifications. Because the grammar, reservoir, and continuous search are finite, the reported winner is the best topology found within each evaluated design space rather than a proof of global optimality.

The simulations and optimization routines were implemented in Python using double-precision numerical linear algebra. A generative artificial intelligence model (OpenAI Codex) was used for code development and execution.

\section{Results}

\subsection{Nonresonant feed-forward spectrometers}

\begin{figure*}[t]
\centering
\includegraphics[width=\linewidth]{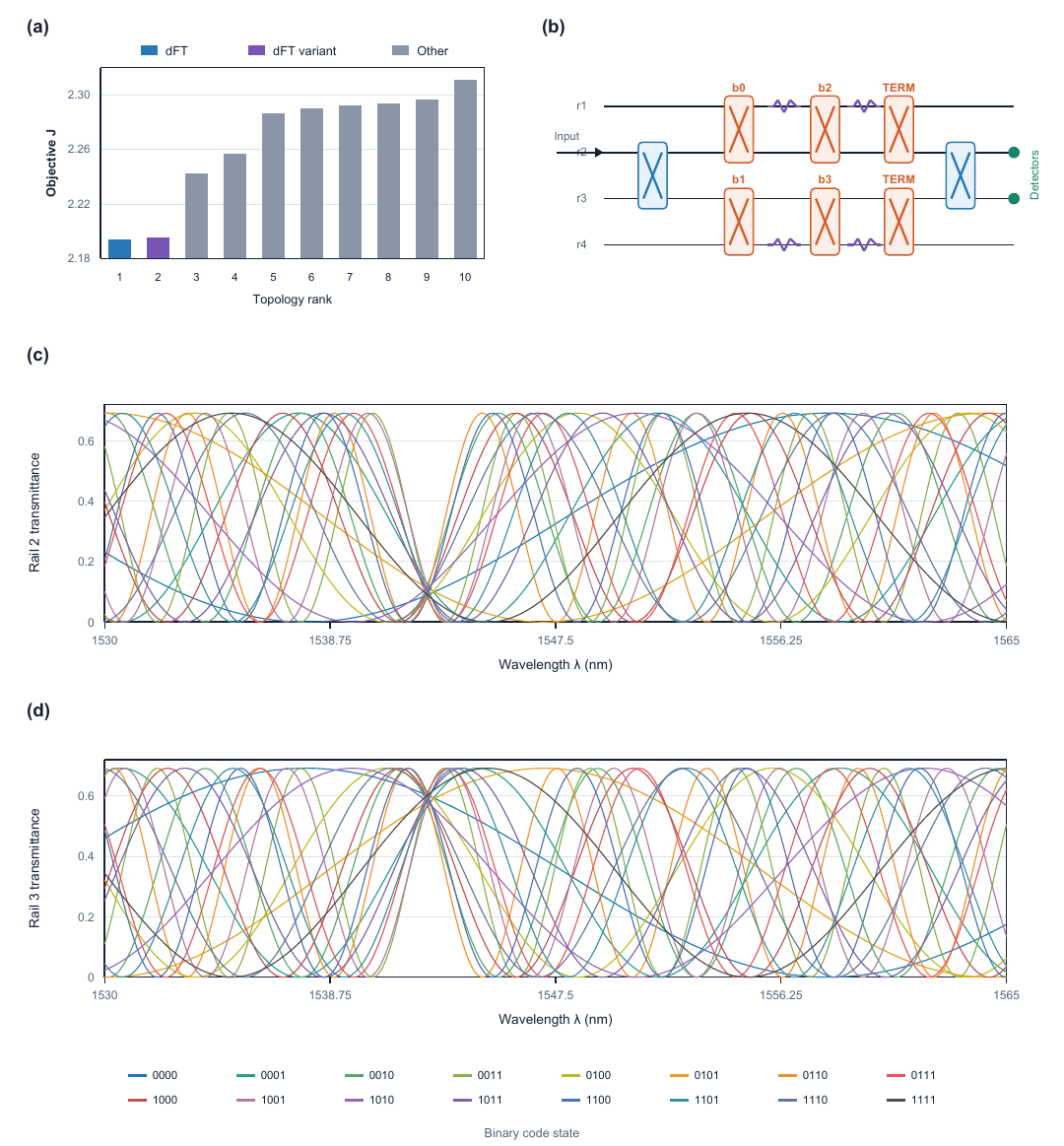}
\caption{Nominal $K=4$ topology search results. (a) Objective values of the ten highest-ranked candidates from one search. Rank 1 is dFT, rank 2 is the dFT variant detailed in Supplementary Fig.~S1(b), and ranks 3--10 are structurally distinct non-dFT circuits. (b) Generic organization of the selected dFT topology. (c,d) The 16 state-resolved transmission spectra at detected rails 2 and 3, respectively. The same color denotes the same four-bit control state in both panels.}
\label{fig:nominal_k4}
\end{figure*}

We first considered the nominal $K=4$ design space, in which four independent binary controls generated 16 optical states. The search assigned dFT the lowest $J$ [Fig.~\ref{fig:nominal_k4}(a,b)]. Two repeat searches selected the same dFT topology with comparable architecture-screening scores albeit having different delay parameters; their detailed numerical results are provided in Supplementary Sec.~S2. These consistent results support the architectural motif while demonstrating that multiple parameter configurations are viable.

The nominal runner-up, with $J=2.195762$, was not a genuinely distinct alternative architecture. It preserved the dFT backbone but replaced one switch-controlled differential-delay stage with a binary PS [Supplementary Fig.~S1(b)]. After this dFT variant, the next candidate had $J=2.242163$, and every other top-ranked non-dFT circuit was likewise appreciably inferior under the screening objective [Fig.~\ref{fig:nominal_k4}(a)]. The topology, grammar word, optimized parameters, and 16-state responses of this highest-ranked true non-dFT circuit are reported in Supplementary Fig.~S2. The true non-dFT candidates interrupted the balanced two-arm, switch-selected delay sequence with additional mixing, asymmetric phase coding, or terminal projection. Within the evaluated component budget, these departures did not create enough independent response diversity to compensate for less uniform power collection and less balanced interferometric coding.

The generic selected circuit is depicted in Fig.~\ref{fig:nominal_k4}(b), and the 16 complementary output responses of the parameter-optimized design are shown in Fig.~\ref{fig:nominal_k4}(c,d). We note that the parameter-optimized designs deviate from the classical dFT recipe: while the two WCs consistently converged near equal splitting, the optimized DL phase delays are not constrained to the canonical power-of-two ratios~\cite{Kita2017Scaling,Kita2018DFT}. In this topology family, balanced splitting and switch-selected delays generate a structured family of Fourier-like responses, while state-programmed terminal routing retains light that would otherwise be discarded. This combination allows dFT to preserve both response diversity and throughput even though the optimized physical delays vary between searches.

Additional nonresonant campaigns, detailed in the Supplementary Information, tested four boundary conditions: doubled component loss at $K=4$; dispersive DL, PS, and OS models over the C band; the same dispersive model over 1500--1600~nm; and a $K=6$ component budget. Canonical dFT was the best topology found in the wide-band and $K=6$ campaigns. In the C-band dispersion campaign, a dFT variant that replaced one OS with a PS was numerically tied with canonical dFT, while the true non-dFT topologies were inferior. The $K=4$ dFT circuit uses four independent OSs distributed across cascaded delay-selection stages, together with terminal-selecting OSs, so increasing switch insertion loss directly penalizes its otherwise favorable balanced encoding. Under doubled loss, a shallower dFT variant that replaces one lossy switch/differential-delay stage with a PS retains more optical power and becomes preferable despite slightly weaker response conditioning. Because the PS remained lossless in this model, this preference is conditional on zero PS insertion loss; nonzero PS loss would reduce the variant's throughput advantage. Supplementary Fig.~S3 shows this topology, its 16-state spectra, and the associated performance tradeoff. The highest-ranked circuit therefore remains within the dFT family; structurally distinct non-dFT topologies were inferior in the reproducible doubled-loss campaign.

\begin{figure*}[p]
\centering
\includegraphics[width=0.98\linewidth]{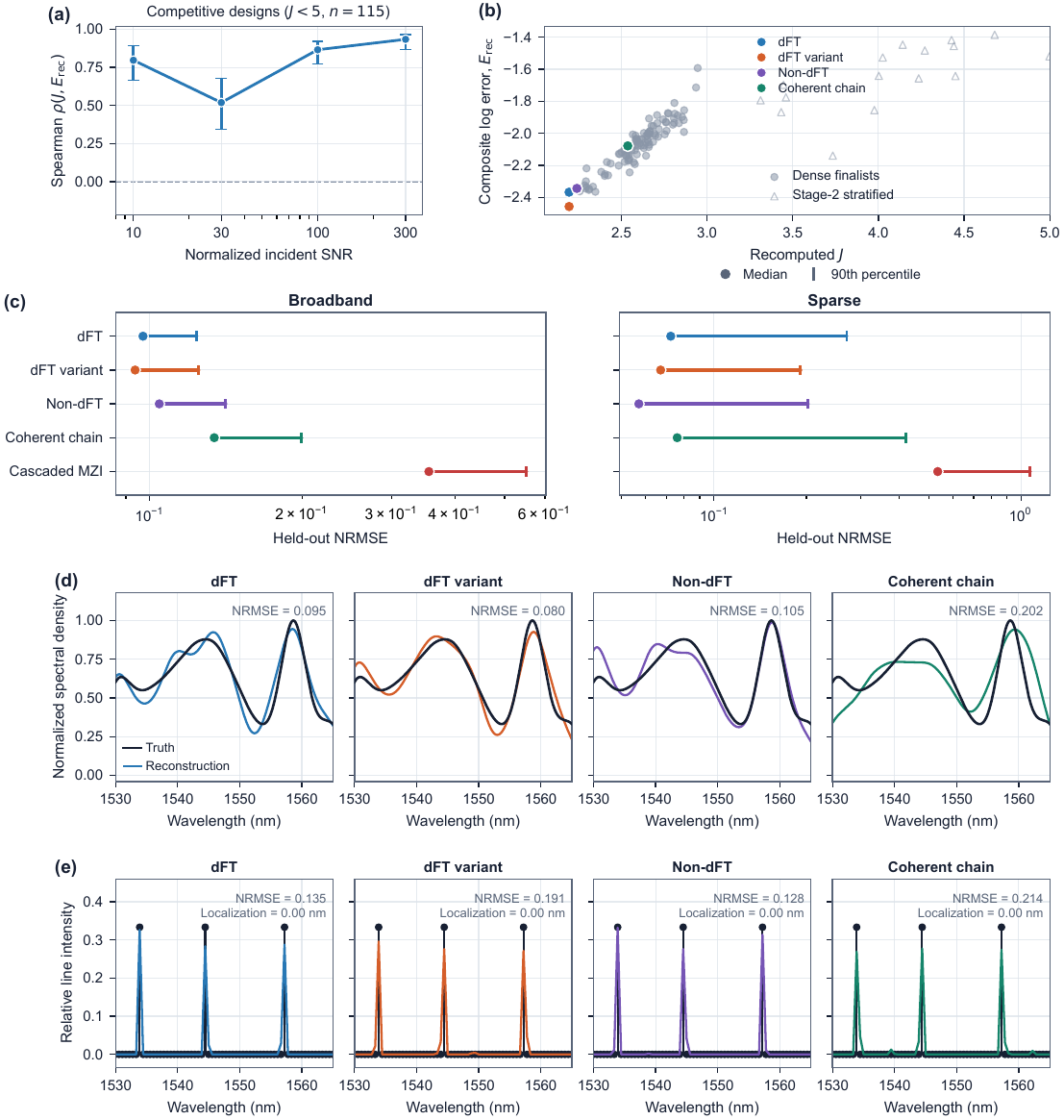}
\caption{Direct reconstruction validation of the architecture-screening objective. (a) Spearman correlation between $J$ and the empirical composite reconstruction error $E_{\mathrm{rec}}$ for the 115 competitive $K=4$ designs with $J<5$; error bars are 95\% intervals from 1,000 bootstrap resamples. (b) $J$ versus $E_{\mathrm{rec}}$ for the same competitive designs at $\mathrm{SNR}_{\mathrm{in}}=300$ (49.5~dB). Filled circles and open triangles distinguish densely evaluated finalists and stage-2 stratified candidates; colored points identify the four reference or finalist designs present in this range. (c) Held-out broadband and sparse reconstruction for five representative $K=4$ architectures at the same SNR. Circles show median NRMSE, horizontal segments extend to the 90th percentile, and terminal ticks mark that percentile. (d) Representative held-out mixed broadband reconstruction for dFT, the dFT variant, the best structurally non-dFT candidate, and the coherent chain. (e) Representative held-out equal-strength three-line reconstruction for the same designs; mean line-localization error is also reported. All designs in each example use the same truth and detector noise realization.}
\label{fig:reconstruction_validation}
\end{figure*}

Because $J$ screens response matrices without selecting a decoder, we next tested its relationship to direct held-out reconstruction [Fig.~\ref{fig:reconstruction_validation}]. Within the 115 competitive designs having $J<5$, its rank association with the empirical composite reconstruction error was nonmonotonic at lower SNR and became strong as detector noise decreased [Fig.~\ref{fig:reconstruction_validation}(a)]. The Spearman correlation was $\rho=0.866$ (95\% bootstrap interval 0.775--0.921) at $\mathrm{SNR}_{\mathrm{in}}=100$ and $\rho=0.934$ (0.870--0.965) at $\mathrm{SNR}_{\mathrm{in}}=300$. The strong correlation between $J$ and the reconstruction error is visualized in Fig.~\ref{fig:reconstruction_validation}(b) for $\mathrm{SNR}_{\mathrm{in}}=300$. The complete cohort, including the reciprocal-singular-value floor regime at large $J$, is shown and analyzed in Supplementary Fig.~S7. These results support $J$ as an effective architecture-screening surrogate for the low-$J$ competitive designs, while showing that it remains distinct from direct, prior- and decoder-dependent reconstruction error.

Direct nonnegative reconstruction of the five representative $K=4$ architectures further resolves differences within the low-$J$ region [Fig.~\ref{fig:reconstruction_validation}(c)]. At $\mathrm{SNR}_{\mathrm{in}}=300$, the dFT variant slightly outperformed canonical dFT in median broadband and sparse NRMSE, demonstrating that near-tied surrogate scores can exchange decoder-specific rankings. The best structurally non-dFT candidate had worse broadband reconstruction but better sparse reconstruction than dFT, whereas the coherent chain developed a substantially larger sparse-error tail and the cascaded MZI was inferior in both tasks. The broadband example in Fig.~\ref{fig:reconstruction_validation}(d) uses the dFT held-out sample closest to its class-specific median NRMSE. For the sparse example in Fig.~\ref{fig:reconstruction_validation}(e), the shared trial was selected by minimizing the summed squared log-distance from the class-specific median NRMSE of each displayed design. Each panel reuses the selected truth and detector noise realization for every design. The examples illustrate both the balanced broadband performance of the dFT family and the task dependence of individual reconstruction outcomes; thus a single example is not used to rank the architectures.

\clearpage

\subsection{Spectrometers incorporating resonators}

We next tested whether local resonant elements supplied a useful coding mechanism beyond the best ring-free circuits. These calculations were bounded augmentation and replacement studies rather than unrestricted searches over resonator-network topology. The catalogues nevertheless cover the principal circuit motifs used in integrated resonator spectrometers: rings coupled to interferometer arms, tunable all-pass and add--drop filters, multiple-ring filter networks, and ring-only chains~\cite{Zheng2019RingAssisted,Zheng2019TunableMicroring,Sun2022FSRFree,Xu2023PhotonicMolecule,Xu2023SingleResonator,Liu2025MicroringNetwork,Yao2024Metrology}.

\subsubsection{Passive rings}

\begin{figure*}[t]
\centering
\includegraphics[width=\linewidth]{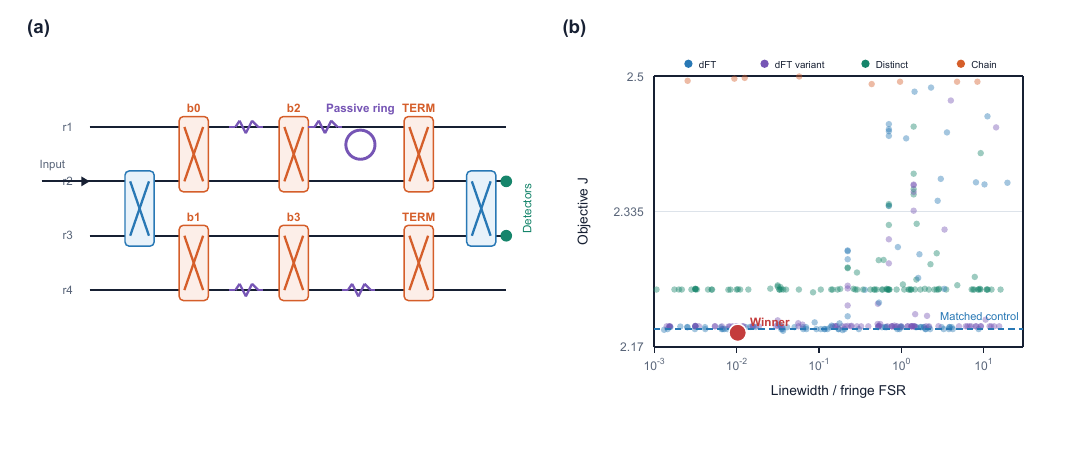}
\caption{Passive-ring results. (a) The best candidate retains the dFT topology and adds one all-pass ring to a switched path. Its ring parameters and comparison metrics are reported in Supplementary Tables~S10 and S12. (b) Objective versus ring linewidth normalized by the conservative interferometric fringe free spectral range, with color identifying the ring-free parent topology. The red point and dashed line identify the selected candidate and the corresponding ring-free circuit, respectively; higher-score candidates are omitted for clarity.}
\label{fig:passive_ring}
\end{figure*}

The best passive-ring candidate retained the dFT topology and added a heavily overcoupled, high-intrinsic-$Q$ ring to one switched path [Fig.~\ref{fig:passive_ring}(a)]. Its optimized device parameters and the metrics of the corresponding ring-free circuit are reported in Supplementary Tables~S10 and S12. This nearly lossless all-pass element introduces only a narrow Fano-like perturbation when embedded in the interferometer; it does not supply a new broadband coding dimension. Removing the ring while retaining the non-ring parameters therefore produced almost no change in throughput or multi-line conditioning, and only a minute change in the aggregate objective.

To compare the ring linewidth with the spectral scale of each parent interferometer, we define the conservative fringe free spectral range as $\Delta\lambda_{\mathrm{fringe}}=\lambda_0^2/(n_gL_{\Sigma})$, where $L_{\Sigma}$ is the sum of all explicitly specified excess DL lengths in the ring-free parent circuit. Because $L_{\Sigma}$ upper-bounds the differential length between any two forward paths, $\Delta\lambda_{\mathrm{fringe}}$ is a lower bound on the actual interferometric fringe free spectral range. In the wider catalogue [Fig.~\ref{fig:passive_ring}(b)], many optimized rings are effectively transparent, so their objectives approach those of the small set of discrete ring-free parent circuits and form horizontal bands. Passive rings can thus fine-tune a dFT response locally, but no resonant topology provided a material advantage commensurate with the added device complexity.

\subsubsection{Active rings}

\begin{figure*}[t]
\centering
\includegraphics[width=\linewidth]{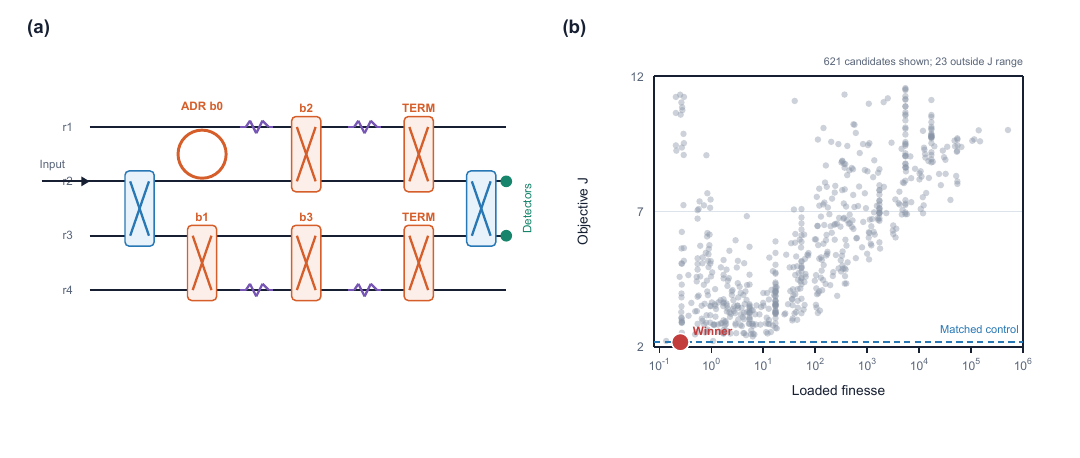}
\caption{Active-ring results. (a) The best candidate replaces one dFT optical switch with an ideal active add--drop ring. The ring parameters, complete non-ring parameter vectors, and comparison metrics are reported in Supplementary Tables~S10--S12. (b) Objective versus loaded finesse for the active-ring catalogue. Gray points are candidates within the displayed objective range, the red point is the refined selected candidate, and the dashed line is the separately reoptimized ring-free circuit; higher-score candidates are omitted for clarity.}
\label{fig:active_ring}
\end{figure*}

The active-ring catalogue produced a nominally larger, but still small, improvement over its separately reoptimized ring-free circuit by replacing one dFT OS with an ideal active add--drop ring (ADR) [Fig.~\ref{fig:active_ring}(a)]. The replacement preserves the dFT organization and improves the modeled throughput and absolute noise-gain term, while the multi-line conditioning changes slightly in the unfavorable direction (Supplementary Table~S12). The complete optimized non-ring parameter vectors of both circuits are reported in Supplementary Table~S11.

The optimized ADR parameters impose an important qualification on this result. The ring is extremely overcoupled, its linewidth exceeds its free spectral range, and its binary tuning swing is close to $\pi$ (Supplementary Table~S10). The selected point consequently lies in the low-finesse regime of the evaluated catalogue [Fig.~\ref{fig:active_ring}(b)]. In this regime the ideal ADR behaves mainly as a low-loss cross connection plus a state-dependent path phase. The implemented symmetric drop-path phase convention allows the optimizer to exploit that phase even when useful resonant amplitude modulation is negligible, and the model assigns no excess loss to the ring couplers.

The concentration of lower $J$ values at smaller loaded finesse in Fig.~\ref{fig:active_ring}(b) is consistent with this cross-plus-phase interpretation. When the linewidth approaches or exceeds the free spectral range, overlapping resonances make the ideal ADR route light between rails over most of the modeled band, while the near-$\pi$ tuning swing supplies a broadband state-dependent phase functionally similar to a PS. Increasing the finesse confines both effects to progressively narrower wavelength intervals, so away from resonance the replacement no longer performs the broadband switching operation required by the dFT backbone. Narrow resonant features therefore do not create enough broadband response diversity to improve the global measurement matrix. Because the plotted candidates also differ in parent topology and ring count, this trend should not be interpreted as a universal monotonic dependence on finesse. The resulting improvement thus results from the near-lossless nature of the ADR, but it is not evidence that an active resonator provides a robust physical advantage over a ring-free switch or that overlapping resonances improve amplitude coding.

\section{Discussion}

The results point to two basic rules for reconstructive spectrometer design. First, the measurement signatures associated with different spectral basis functions should be as disparate as possible, giving a response matrix with low mutual correlation and well-conditioned singular directions. Second, those signatures should be generated with high optical throughput: useful light should reach the detectors rather than be attenuated by lossy elements, rejected at dump ports, or dissipated in resonators. In an ideal lossless, detector-noise-limited system, a limiting target that combines these rules is a balanced binary measurement code. Let $\mathbf{S}$ contain only 0 and 1, with each entry specifying whether a spectral channel is routed to a detector in a given measurement. The entries alone do not guarantee optimality; the patterns must be chosen so that the centered matrix $\mathbf{H}=2\mathbf{S}-\mathbf{U}$, where $\mathbf{U}$ is the all-ones matrix, has orthogonal columns, $\mathbf{H}^{\mathsf{T}}\mathbf{H}=M\mathbf{I}$ for $M$ measurements. This is the unipolar form of a Hadamard code; the closely related odd-order construction is conventionally termed a Hadamard $S$-matrix~\cite{Harwit1979Hadamard}. With complementary dual-output detection, $\mathbf{S}$ and $\mathbf{U}-\mathbf{S}$ can be recorded simultaneously and subtracted to recover the signed Hadamard code, while every wavelength is directed to one of the two detectors rather than discarded.

Such an ideal binary code is difficult to realize as a compact broadband photonic circuit. Its rectangular pass--stop response functions contain many spectral Fourier components; synthesizing their abrupt transitions generally requires many interferometric delays or a high-order resonant filter. The added circuit depth, couplers, and resonators then introduce precisely the insertion loss and unused ports that the throughput rule seeks to avoid, while also increasing footprint, calibration burden, and sensitivity to fabrication error. The practically preferred topology must therefore trade the mathematical distinctiveness of the measurement functions against the physical complexity required to synthesize them.

dFT emerged as the best compromise within the evaluated design spaces. A two-arm interferometer produces complementary responses of the form $[1\pm\cos\phi(\lambda)]/2$; when $\phi$ is set by a fixed path delay, the wavelength-dependent term is a single harmonic in optical wavenumber. Switch-selected differential delays vary that component across the binary states without requiring a high-order filter. Near-balanced WCs maintain interference contrast, the two detected outputs preserve complementary information, and state-programmed terminal routing recovers light from the outer rails instead of sending it directly to a dump. The resulting smooth Fourier-like functions remain sufficiently distinct for stable reconstruction while being generated by a shallow circuit. This physical organization, rather than a unique delay recipe, is the recurring result: the optimized delays did not generally follow the canonical power-of-two prescription~\cite{Kita2017Scaling,Kita2018DFT}.

The doubled-loss study clarifies that the preferred realization within the dFT family is conditional on the device technology. A dFT circuit obtains its coding diversity through several cascaded OSs, including terminal-selecting switching elements; when the loss assigned to each switch is doubled, that depth imposes a substantial throughput penalty. The best circuit therefore becomes a shallower dFT variant in which a PS replaces one lossy switch-and-differential-delay stage. This substitution improves throughput enough to outweigh the modest reduction in response conditioning, while preserving the Fourier-like interferometric backbone that distinguishes the dFT family. The result is thus an intra-family transition rather than evidence that a structurally distinct non-dFT topology becomes preferable. It is conditional on the assumed zero insertion loss of the PS; a lossy PS would diminish the throughput gained by replacing the OS stage. Consequently, topology and component implementation should not be optimized independently: even when the preferred architectural motif is unchanged, its optimal realization can depend on switch loss, coupler loss, and routing depth.

This technology dependence also illustrates why an inference-oriented architecture surrogate is useful. A correlation-function width measures response similarity after normalization and therefore cannot register a uniform loss of signal. By retaining absolute response amplitude in $G_L$ and supplementing it with the three-line conditioning statistic $C_{3,90}$, $J$ screens both detector-noise amplification and spectral ambiguity~\cite{Ma2026RobustInference}. It therefore makes the two design rules explicit: spectral diversity is rewarded only when it is obtained without an excessive penalty in collected power. Its decoder independence is useful during a large topology search, but also sets an evidence boundary: direct reconstruction remains necessary when task priors, regularization, and finite-SNR behavior matter.

The held-out validation quantifies that boundary. The correlation between $J$ and empirical reconstruction error becomes strong at high SNR but is weaker at lower SNR, where decoder bias and noise floor effects can dominate architectural conditioning. Moreover, the dFT variant slightly surpasses canonical dFT under the tested nonnegative decoders, and the structurally non-dFT finalist trades worse broadband performance for better sparse performance. Thus, $J$ should be used to identify a competitive architecture region rather than to assert a decoder-independent ordering among near ties. The direct tests nevertheless show that dFT and its variants provide the most favorable overall tradeoff across the broadband and sparse tasks, consistent with the physical design rules.

The ring catalogues reinforce this interpretation. Adding the optimized passive ring to dFT produced only a minute local change: the ring acted as an almost lossless all-pass perturbation rather than a new broadband coding degree of freedom. The active-ring result was likewise not evidence for superior resonant amplitude coding. Its optimized ideal ADR was extremely overcoupled and operated in a low-finesse regime, where it functioned mainly as a low-loss cross connection with a state-dependent phase. The small modeled improvement therefore arose from the effective routing and phase operation, together with the assumed absence of excess coupler loss, rather than from spectrally selective resonance. Realistic active-ring performance will depend on coupler loss, tuning geometry, thermal crosstalk, and the phase convention at the circuit reference planes. Within the present model, resonators can fine-tune an already competitive topology, but they do not displace its underlying balanced interferometric organization.

Beyond the specific ranking, the grammar provides a systematic way to turn these observations into technology-specific design decisions. Known dFT, cascaded-MZI, phase-chain, and hybrid circuits arise from the same representation and are evaluated by the same scattering-matrix model and reconstruction objective. The search can therefore compare named architectures and previously unnamed variants without assigning either group a privileged physical model. Its output is not only a winner but also a set of conditional design rules: favor balanced splitting and complementary collection; use controllable differential delays to generate nonredundant states; preserve power at terminal routing; optimize delays rather than imposing canonical ratios; and limit lossy circuit depth. Replacing the component models and bounds with measured foundry data would allow the same workflow to identify where these rules change for a particular fabrication platform.

Several limitations define the scope of these conclusions. The grammar is restricted to four feed-forward rails, adjacent couplers, fixed input and detector ports, finite component counts, and bounded passive and active ring catalogues. The optimization uses finite topology reservoirs and parameter budgets, so it cannot prove a global optimum or exclude a missed circuit or continuous solution. The $K=6$ result is one finite-search validation and does not establish asymptotic scaling with $K$; likewise, numerical objectives cannot be compared directly between $K=4$ and $K=6$ or between different analysis bandwidths. The direct-reconstruction tests use synthetic spectra, additive signal-independent noise, and two validation-tuned decoder families; they do not establish task-independent reconstruction superiority. A focused $K=4$ mismatch audit added finite switch extinction, coupler and delay/phase perturbations, nonzero PS loss, and post-calibration drift; a dFT or dFT variant retained lower $J$ and lower median broadband reconstruction error than the structurally non-dFT finalist in every moderate and stress-test realization (Supplementary Sec.~S10). This limited audit does not replace an experiment or a foundry-specific yield analysis. The physical model still neglects polarization, nonlinearities, thermal and electrical power, correlated or signal-dependent noise, and broader calibration mismatch. Future work should incorporate measured component S-matrices and uncertainty distributions, extend the grammar to additional rails, nonadjacent and multimode elements, and more general resonator networks, and experimentally test the predicted topology family using measured calibration responses and application-specific reconstruction tasks. Footprint, tuning energy, calibration time, and fabrication yield can also be introduced as explicit objectives or constraints.

\section{Conclusion}

We developed a grammar-based framework that searches circuit connectivity and continuous parameters together for reconfigurable integrated photonic spectrometers. The grammar composes standard photonic building blocks, canonicalizes equivalent circuits, removes physically inactive or redundant candidates, and evaluates the surviving topologies using a common scattering-matrix model. Coupling this search to the decoder-independent $J$ surrogate makes architecture screening sensitive to both noise amplification and optical throughput rather than to spectral correlation width alone. Direct held-out reconstruction confirms strong rank predictiveness among competitive designs while showing that decoder-specific rankings can exchange among near-tied designs.

Across the design spaces evaluated, dFT and its variants consistently delivered the strongest combination of reconstruction robustness and optical throughput. Their recurrence under repeated nominal searches, expanded bandwidth, a larger switch count, and direct broadband--sparse reconstruction identifies balanced Fourier-like coding as a robust architectural motif. In the doubled-loss model, which retained zero PS insertion loss, increased WC and OS loss favored a shallower dFT variant rather than a transition to a true non-dFT topology, while passive and ideal active rings offered no robust advantage attributable to resonant amplitude coding. These results do not establish a universal optimum; instead, they show that spectrometer topology can be selected through a reproducible, physics-informed comparison tied to a specified component library, noise model, design budget, and reconstruction task. The framework provides a route for benchmarking existing architectures and discovering technology-specific variants as integrated photonic platforms and system requirements evolve. More broadly, the same grammar-based circuit topology search can identify task-optimized PIC architectures across sensing, communications, signal processing, and computing. Combined with the component-level inverse design commonly described as 'topology optimization', it offers a path toward an end-to-end PIC design framework spanning circuit architecture and device geometry, with broad relevance to integrated photonic design.

\section*{Funding}

National Institutes of Health (NIH; R01AG077016).

\section*{Disclosures}

The author declares no conflicts of interest.

\section*{Data Availability}

Source codes supporting the simulations in this paper are available at \url{https://github.com/hujuejun/grammar-pic-spectrometer}. The topology searches are stochastic: the released code and settings support independent method-equivalent reruns, but regenerated candidate reservoirs and detailed rankings may differ slightly from the reported searches.

\bibliography{references}

\end{document}